\documentclass[journal,transmag]{IEEEtran}

\usepackage{cite}
\usepackage{graphicx}
\usepackage{epstopdf}
\usepackage{amsthm}
\usepackage[cmex10]{amsmath}
\usepackage{array}
\usepackage{algorithm}
\usepackage{algorithmic}
\usepackage{float}
\usepackage{graphicx}
\usepackage{amssymb}
\usepackage{microtype}
\usepackage{color}
\usepackage{relsize}
\usepackage[font=footnotesize]{caption}
\usepackage{multirow}
\usepackage{balance}
\usepackage{amsmath}

\newcommand{\comment}[1]{}
\bstctlcite{IEEEexample:BSTcontrol}

\newtheorem{theorem}{Theorem}

\newtheorem{definition}{Definition}

\newtheorem{example}{Example}
\newtheorem{claim}{Claim}

\begin{document}

\title{\vspace{-0.2cm}Coding for Channels with SNR Variation: Spatial Coupling and Efficient Interleaving\vspace{-0.3cm}}

\author{\IEEEauthorblockN{Homa Esfahanizadeh$^1$, \textit{Student Member, IEEE}, Ahmed Hareedy$^1$, \textit{Student Member, IEEE},\\ Ruiyi Wu$^1$, Rick Galbraith$^2$, and Lara Dolecek$^1$, \textit{Senior Member, IEEE}}
\IEEEauthorblockA{$^1$ Department of Electrical Engineering, University of California, Los Angeles, Los Angeles, CA 90095 USA\\
$^2$ Hitachi Global Storage Technologies, a Western Digital Company, Rochester, MN 55901 USA\vspace{-0.3cm}
%hesfahanizadeh@ucla.edu, ahareedy@ucla.edu, ruiyiwu@ucla.edu, and dolecek@ee.ucla.edu
}}

\IEEEtitleabstractindextext{
\begin{abstract}
In magnetic-recording systems, consecutive sections experience different signal to noise ratios (SNRs). To perform error correction over these systems, one approach is to use an individual block code for each section. However, the performance over a section affected by a lower SNR is weaker compared to the performance over a section affected by a higher SNR. A commonly used technique is to perform interleaving across blocks to alleviate negative effects of varying SNR. However, this scheme is typically costly to implement and does not result in the best performance.
Spatially-coupled (SC) codes are a family of graph-based codes with capacity approaching performance and low latency decoding. An SC code is constructed by partitioning an underlying block code to several component matrices, and coupling copies of the component matrices together.
%Here, each SC code spans more than one section. 
The contribution of this paper is threefold. First, we present a new partitioning technique to efficiently construct SC codes with column weights 4 and 6. Second, we present an SC code construction for channels with SNR variation. Our SC code construction provides local error correction for each section by means of the underlying codes that cover one section each, and simultaneously, an added level of error correction by means of coupling among the underlying codes. Consequently, and because of the structure of SC codes, more reliable sections can help unreliable ones to achieve an improved performance. Third, we introduce a low-complexity interleaving scheme specific to SC codes that further improves their performance over channels with SNR variation.
%For partitioning block codes, we use the minimum overlap (MO) partitioning technique to reduce the population of problematic objects. Moreover, we extend the MO partitioning technique to construct SC codes with higher column weights. 
Our simulation results show that our SC codes outperform individual block codes by more than 1 and 2 orders of magnitudes in the error floor region compared to the block codes with and without regular interleaving, respectively. This improvement is more pronounced by increasing the memory and column weight.
\end{abstract}

\vspace{-0.2cm}
\begin{IEEEkeywords}
Error floor, LDPC codes, magnetic-recording, MO partitioning, SNR variation, spatially-coupled codes.\vspace{-0.3cm}
\end{IEEEkeywords}}

\maketitle

\section{Introduction}

\IEEEPARstart{M}{odern} magnetic recording (MR) systems require error correcting codes (ECCs) with outstanding error floor performance. Low-density parity-check (LDPC) codes are a primary choice for MR systems because of their error correcting capabilities {\cite{6253209,5402492,4100818}}. In a magnetic-recording device, some sections can be more error prone than other sections because of the read/write mechanism and physical properties of the device \cite{varnica2012interleaved}. A realistic channel model for magnetic recording systems must consider the variation of signal to noise ratio (SNR) among consecutive sections of a hard disk drive.

In this paper, we develop ECCs that address the SNR variation for data storage systems. For channels with uniform SNR, i.e., channels with a single SNR, the goal is to find a code that achieves a certain level of bit error rate (BER) for that SNR. For channels with SNR variation, conventional ECCs are designed to achieve the target BER for the section with the lowest SNR. For the sections with higher SNRs, this approach results in an additional redundancy which is not required to achieve the target BER.

One solution for handling the non-uniformity of SNR is using interleaving. In this approach, we assume that the channel consists of $N$ sections of equal sizes, and the length of each codeword is also equal to the length of a section. Each codeword is divided into $N$ chunks. Then, different chunks from all codewords are interleaved such that one chunk from each codeword is passed through one section of the channel. As a result, the average SNR that all codewords are affected by is the same and equal to the average SNR of the channel, and one just needs to consider the average SNR rather than the worst SNR in the code design process \cite{varnica2012interleaved}. One disadvantage of this approach is that the length of each codeword is equal to the length of one section of the channel which might be relatively short.

The authors in \cite{7117410} have recently presented a concatenated code design scheme for magnetic recording systems that considers the SNR variation. Their approach consists of a number of inner codes and one outer code. While the inner codes span one section each, the outer code spans many sections, possibly an entire track.

Spatially-coupled (SC) codes are a family of graph-based codes that have attracted significant attention because of their capacity approaching performance and low decoding latency \cite{782171,6374679}. SC codes are constructed by partitioning a parity-check matrix $\bold{H}$ of the underlying block code into component matrices $\{\bold{H}_0, \bold{H}_1,\cdots, \bold{H}_m\}$, $\bold{H}=\sum_{y=0}^{m}\bold{H}_y$, and coupling $L$ replicas of the component matrices together to obtain the parity-check matrix $\bold{H}_\text{SC}$, see Fig.~1. The parameters $m$ and $L$ are known as the memory and coupling length, respectively. %By using window decoder for SC codes, the decoding latency only depends on the length of the underlying block code and the size of the sliding window which is much lower than the latency of a block code with similar length \cite{6374679}.

\begin{figure}
\centering
\includegraphics[width=0.27\textwidth]{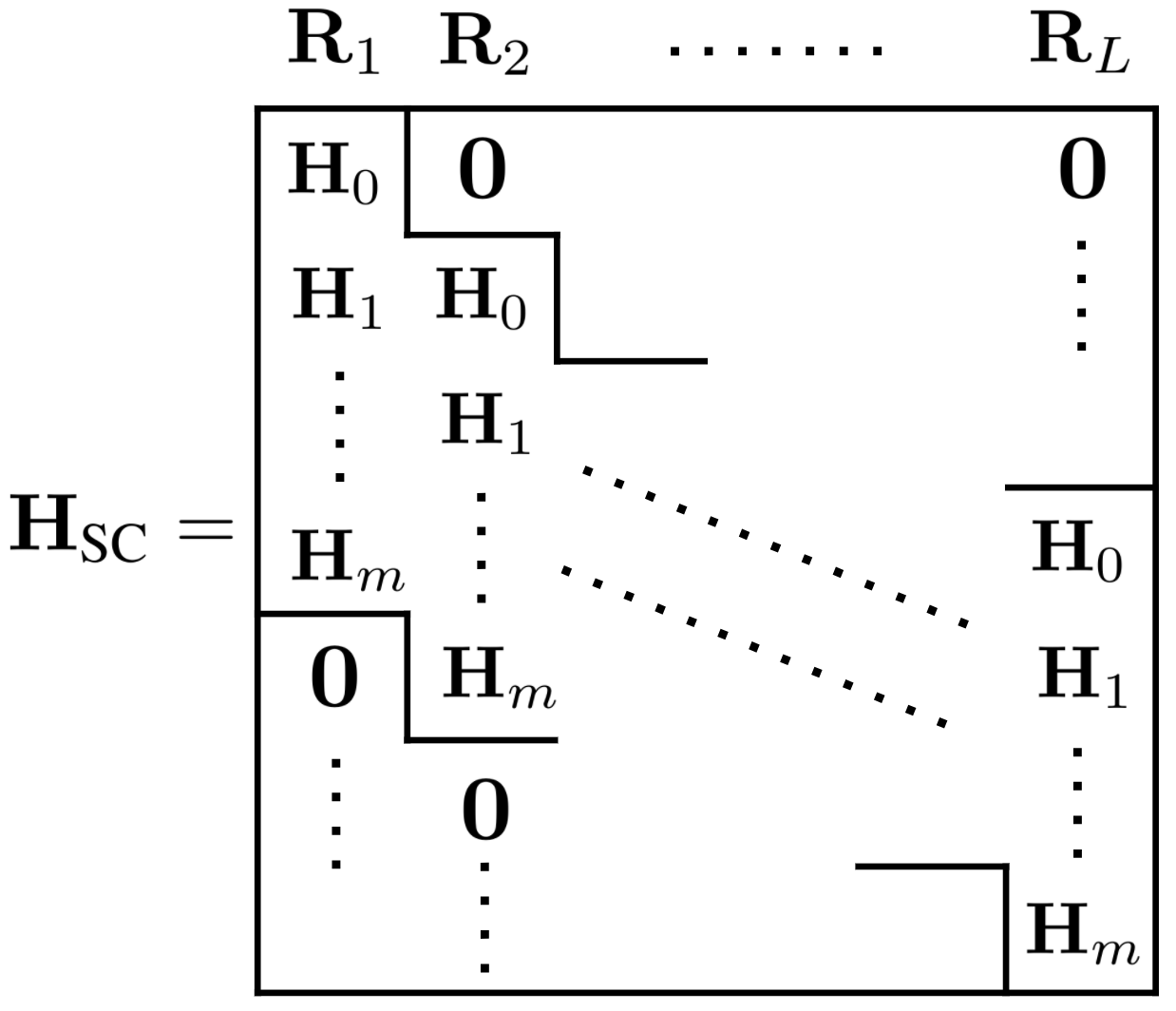}\\
\caption{The parity-check matrix of an SC code with memory $m$ and coupling length $L$.}
\vspace{-0.5 cm}
\end{figure}

In \cite{7922520,7439792,6883627}, SC codes over non-uniform channels are studied. In these papers, the asymptotic performance of SC codes over burst erasure channels is considered. In this paper, we present \textit{a finite-length analysis and construction of SC codes for channels with SNR variation}. We use circulant-based (CB) LDPC codes as the underlying block codes \cite{1362891} and the minimum overlap (MO) technique for the partitioning \cite{MOP_ISIT_2017}. The MO partitioning technique was introduced for partitioning CB block codes and constructing SC codes with column weight $\gamma=3$ . The MO technique aims to minimize the population of problematic combinatorial objects inside each component matrix to achieve a good (error floor) performance. In this paper, we use the MO approach to construct SC codes with $\gamma=3$. Moreover, we further extend this approach to construct SC codes with $\gamma=2(m+1)$. We also apply a circulant power optimizer to further reduce the number of problematic objects in the final SC code by heuristically adjusting the circulant powers \cite{2018arXiv180206481E}.

The contribution of this paper is threefold:
1) we extend MO partitioning to construct SC codes with higher column weights,
2) we show that, compared to uncoupled block codes, SC codes show a superior performance over channels with SNR variation
thanks to their intrinsic structure, and we demonstrate that SC codes with higher memories are more robust against the SNR variation,
and 3) we introduce an efficient interleaving scheme specific to SC codes over channels with SNR variation that further improves their performance. The proposed interleaving has a lower complexity compared to the regular interleaving, and it ensures that each check node (CN) in the structure of SC codes is connected to variable nodes (VNs) with different reliabilities.

According to our simulation results, for column weight $3$ and compared to the scheme of block codes with interleaving, our SC code design achieves over $1$ order of magnitude performance improvement in the error floor region for the case of memory $1$ and more improvement for higher memories. We also demonstrate the good performance of SC codes with higher column weight constructed by our new partitioning scheme. Our simulation results verify that the performances of our presented SC codes over channels with SNR variation are close to their performance over uniform channels with the same average SNR, and the introduced efficient interleaving technique notably reduces this gap in the performance. %Besides, we compare the performance of block and SC codes, and demonstrate that the memory of SC codes make them more robust against the SNR variation.

The rest of this paper is organized as follows: In Section II, we review the necessary background. In Section III, we present our extension of the MO technique for partitioning block codes and constructing SC codes with $\gamma=2(m+1)$. In Section IV, we present our SC code construction and the low-complexity interleaving scheme for SC codes to compensate for the negative impact of SNR variation. The simulation results are given in Section V, and conclusions appear in Section VI.\vspace{-0.1cm}

\section{Preliminaries}

In this section, we first describe circulant-based SC codes. Next, we revisit the combinatorial problematic objects in the error floor region. Then, we review the system model that we use to describe a channel with SNR variation. Finally, we review the regular interleaving technique for channels with SNR variation.
\vspace{-0.15cm}
\subsection{Circulant-Based LDPC Codes and SC Codes}

Regular CB codes are a class of structured $(\gamma,\kappa)$ LDPC codes, where $\gamma$ is the column weight of the parity-check matrix (variable node degree in the Tanner graph), and $\kappa$ is the row weight (check node degree). CB codes can offer simple hardware implementation thanks to their structure {\cite{1362891}}. Let $\bold{H}$ be the parity-check matrix of the underlying CB code, $\bold{H}$ consists of $\gamma \kappa$ circulants. Each circulant is of the form $\sigma^{f_{i,j}}$, where $0\leq i \leq \gamma -1$, $0\leq j \leq \kappa -1$, and $\sigma$ is the $z\times z$ identity matrix shifted one unit to the left. %For example, Array-based (AB) codes are CB codes with $f_{i,j}=ij$, $\kappa=p$, and girth $6$ \cite{ABCodesRef,5361488}.

SC codes have parity-check matrices with a band-diagonal structure. An SC code is constructed as follows. First, $\bold{H}$ is partitioned into ($m+1$) disjoint component matrices (of the same size as $\bold{H}$): $\bold{H}_0$, $\cdots$, $\bold{H}_m$, where $m$ is the memory. Each component matrix $\bold{H}_y$, $0\leq y \leq m$, contains some of the circulants in $\bold{H}$ and zeros elsewhere such that $\bold{H}=\sum_{y=0}^{m}\bold{H}_y$. Second, $L$ copies of component matrices are pieced together to obtain the parity-check matrix $\bold{H}_\text{SC}$ of the SC code, where  $L$ is the coupling length, see Fig.~1. A replica is a collection of columns of $\bold{H}_\text{SC}$ that contains one submatrix $[\bold{H}_0^\textnormal{T}\cdots \bold{H}_m^\textnormal{T}]^\textnormal{T}$ and zero elsewhere. An SC code with coupling length $L$ has $L$ replicas, $\{\bold{R}_1,\cdots,\bold{R}_L\}$, see Fig.~1. {The {design} rate of an SC code with parameters $\kappa$, $\gamma$, $m$, and $L$ is $r_\textnormal{d}=1-\frac{(L+m)\gamma}{L\kappa}$, which is close to the {design} rate of its underlying block code, $r_\textnormal{d}=1-\frac{\gamma}{\kappa}$, for large values of $L$.}

A recently introduced approach for partitioning the underlying block code is MO partitioning \cite{MOP_ISIT_2017}. In this approach, the matrix $\bold{H}$ is partitioned into several component matrices such that the overlap of each pair of rows of circulants in each component matrix is minimized. The MO partitioning assumes a balanced distribution of circulants among component matrices. The MO partitioning significantly outperforms the previously well-known cutting vector partitioning \cite{6874960,7529198}. The MO partitioning is originally presented to construct SC codes with $\gamma=3$. In Section~III, we extend the MO partitioning to construct SC codes with higher column weights.

The protograph matrix of a code is obtained by replacing each non-zero circulant in the parity-check matrix with scalar $1$ and each zero circulant with scalar $0$, and it is denoted as $\bold{H}_\text{SC}^\textnormal{p}$ (or $\bold{H}^\textnormal{p}$ for the block code). The MO partitioning aims at constructing the protograph matrix of an SC code with a reduced number of problematic objects. Having the matrix $\bold{H}_\text{SC}^\textnormal{p}$, the next stage is lifting to construct $\bold{H}_\text{SC}$, where we apply a program introduced in \cite{2018arXiv180206481E} to optimize the powers of non-zero circulants in the final matrix. The program is called the circulant power optimizer (CPO), and it starts with an initial set of circulant powers and adjusts the powers heuristically to further reduce the number problematic objects.
%In this paper, the circulant powers of an SC code is shown by a $\gamma\times\kappa$ matrix with entries in $\{0,\cdots,p-1\}$. Each entry shows the power of the corresponding circulant in the underlying block code.
\vspace{-0.3cm}
\subsection{Combinatorial Objects of Interest}

Under iterative decoding, certain combinatorial objects in the graph of LDPC codes cause the error floor phenomenon \cite{Richardson2003,5361488}. We recall the definition of absorbing sets (ASs) and trapping sets (TSs) as the primary objects of interest here. The elementary ASs are problematic over AWGN channels \cite{5361488}, and we are interested in elementary ASs here because each section of our model for a channel with SNR variation is an AWGN channel. For the case of non-AWGN sections, one can refer to the problematic objects for non-canonical channels, e.g., \cite{7482740,7553518}. The common subgraph of several dominant ASs for a code might appear as a TS.

\begin{definition}
(cf. \cite{Richardson2003}) Consider a subgraph induced by a subset $\mathcal{V}$ of VNs in the graph of a binary LDPC code. Let $\mathcal{E}$ be the set of even degree CNs connected to $\mathcal{V}$, and similarly, let $\mathcal{O}$ be the set of odd degree CNs connected to $\mathcal{V}$. The set $\mathcal{V}$ is said to be an $(a,b)$ TS if the size of $\mathcal{V}$ is $a$ and the size of $\mathcal{O}$ is $b$.
\end{definition}

\begin{definition}
(cf. \cite{5361488}) An $(a,b)$ AS is a TS with subset $\mathcal{V}$ of VNs such that each VN in $\mathcal{V}$ is connected to strictly more CNs in $\mathcal{E}$ rather than $\mathcal{O}$. An AS is elementary if all CNs have degree one or two.
\end{definition}

For AWGN channels, the $(3,3(\gamma-2))$ AS/TS is the ommon substructure of interest since it not only appears frequently in the error profile of graph-based codes with $\gamma=3$, but it also appears as a substructure of many larger ASs for codes with $\gamma>3$, e.g., \cite{7529198,2018arXiv180206481E}. The configuration of a $(3,3)$ AS is depicted in Fig.~2.\vspace{-0.1cm}

\begin{figure}
\centering
\includegraphics[width=0.115\textwidth]{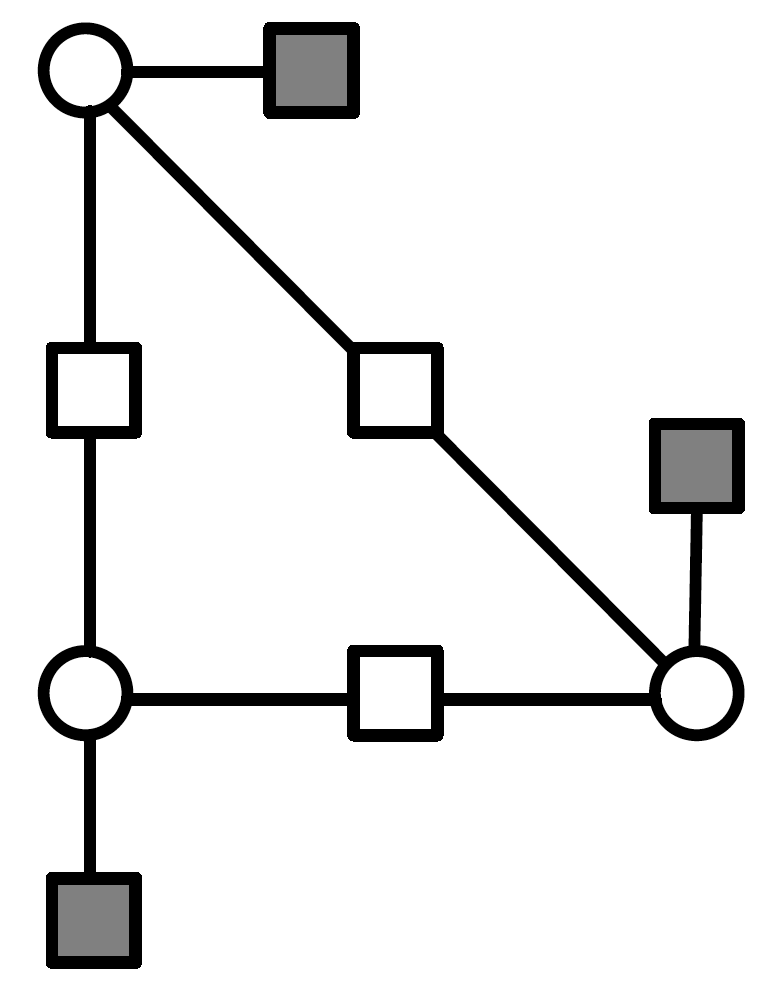}\\
\caption{Graphical representation of $(3,3)$ AS; circles and squares correspond to VNs and CNs, respectively.}
\vspace{-0.3cm}
\end{figure}

\subsection{Channels with SNR Variation}

{Our model for a channel with SNR variation is depicted in Fig.~3(a). It shows a channel with $N$ sections, and each section is considered as an AWGN channel with $\text{SNR}_s$ ($s$ is the section index). For the $s$'th section, we state the SNR as ${(\text{SNR}_s)}_\text{dB}={(\text{SNR}_{\text{abs}})}_\text{dB}+{(\Delta \text{SNR}_{s})}_\text{dB}$, where $\text{SNR}_{\text{abs}}$ is the absolute SNR, $\Delta \text{SNR}_{s}$ is the variation from the absolute SNR for the $s$'th section, and $\text{X}_\text{dB}=10\log_{10}\text{X}$.} We assume all sections have the same length. %In a practical channel model, the values of $\Delta \text{SNR}_{i}$ for different sections are dependent
\vspace{-0.25cm}
\subsection{Interleaving to Mitigate Non-Uniformity}
Based on the described model for the SNR variation, some sections of the channel have higher SNRs than $\text{SNR}_{\text{abs}}$ while others do not. The interleaving technique is used to minimize the negative impact of SNR variation by introducing diversity \cite{varnica2012interleaved}. The idea of interleaving is illustrated in Fig.~3(b). Interleaving is performed on the sequence of codewords before they pass through the channel, and de-interleaving is performed on the received sequence of data and before decoding. 

First, a sequence of data is split into $N$ parts. Each part is coded individually by the same block code to generate $N$ codewords. Next, each codeword is partitioned into $N$ chunks, and different chunks from all codewords are interleaved such that each $N$ consecutive chunks belong to $N$ different codewords. Finally, each $N$ consecutive chunks pass through a separate section of the channel. The interleaving helps to achieve a better error correction capability since the average SNR that a codeword is affected by is the same for all codewords and it is equal to the average SNR of the channel, so each codeword has some reliable bits that can help to decode the less reliable bits.

\begin{figure}
\centering
\includegraphics[width=0.38\textwidth]{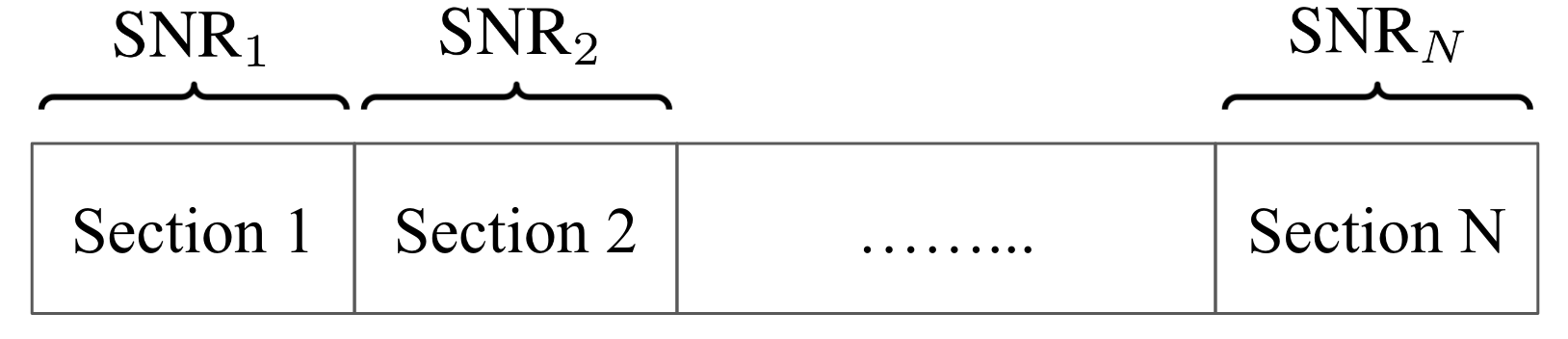}\\
(a)\\
\includegraphics[width=0.37\textwidth]{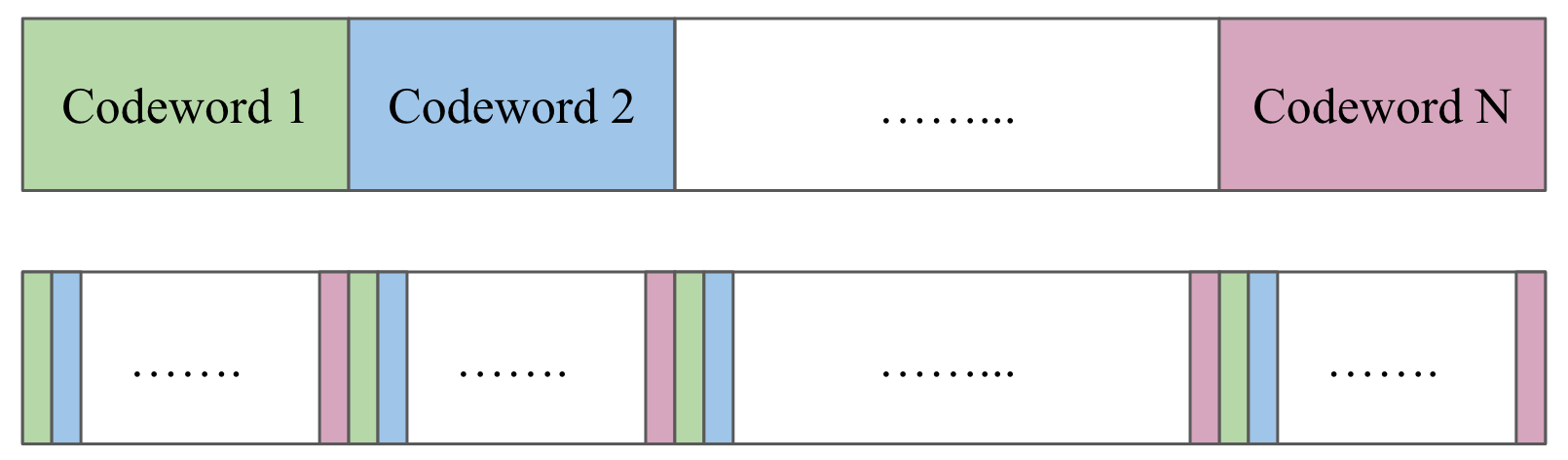}\\
(b)
\caption{(a) A non-uniform channel with $N$ sections. (b) Original and interleaved sequence of encoded data; each color corresponds to one codeword.}
\vspace{-0.4 cm}
\end{figure}

\section{Minimum Overlap Partitioning for Higher Column Weights}
The protograph matrix of an SC code with parameters $\kappa$, $\gamma$, $m$, and $L$ is constructed by partitioning the $\kappa\gamma$ elements, all with values $1$, in the protograph matrix of the underlying code $\bold{H}^\textnormal{p}$ into ($m+1$) component matrices, and piecing $L$ copies of the component matrices together to obtain $\bold{H}_\text{SC}^\textnormal{p}$, see Fig.~1. The partitioning of the underlying block code can be described by a \textit{partitioning matrix} $\bold{PM}=[h_{i,j}]$ of size ${\kappa\times\gamma}$. Each element $h_{i,j}\in\{0,\cdots,m\}$ implies that the corresponding element in $\bold{H}^\textnormal{p}$ (resp., the corresponding circulant in $\bold{H}$) is assigned to $\bold{H}_{h_{i,j}}^\textnormal{p}$ (resp., $\bold{H}_{h_{i,j}}$). 
\begin{example}
A partitioning matrix for an SC code with parameters $\gamma=3$, $\kappa=7$, and $m=1$ is shown below:
\begin{equation*}
\bold{PM}=\begin{tabular}{|c|c|c|c|c|c|c|}
\hline
$0$&$0$&$0$&$0$&$1$&$1$&$1$\\
\hline
$1$&$1$&$1$&$0$&$0$&$0$&$1$\\
\hline
$0$&$1$&$1$&$1$&$0$&$1$&$0$\\
\hline
\end{tabular}.
\end{equation*}
\end{example}

A systematic partitioning scheme, namely Minimum Overlap (MO) partitioning, was recently introduced in \cite{MOP_ISIT_2017} to construct SC codes with a superior performance. The overlap between two rows is defined as the number of columns in which the two rows have non-zero values, simultaneously. 
Let $t_y$ be the maximum overlap between rows (in pairs) of the $y$'th component matrix of the protograph matrix $\bold{H}^\textnormal{p}$, i.e., $\bold{H}_y^\textnormal{p}$.

The technique aims to minimize the overall overlap $t=\min_{y\in\{0,\cdots,m\}}t_y$ in a balanced partitioning in order to prevent certain detrimental structures in the graph representation of the code that are intrinsically formed by the overlaps, e.g., cycles and ASs.

The MO technique in \cite{MOP_ISIT_2017} is presented only for partitioning the codes with column weight $\gamma=3$. Here, we extend the MO technique for partitioning block codes with column weight $\gamma=2(m+1)$ into component matrices with minimum overlap, such that there are two elements (resp., circulants) in each column of $\bold{H}^\textnormal{p}$ (resp., $\bold{H}$) that are assigned to the same component matrix (balanced partitioning). This extension is helpful to systematically construct SC codes with $(\gamma=4,m=1)$ and $(\gamma=6,m=2)$. Theorem~1 states the minimum overlap value, and it also suggests a partitioning construction that achieves the minimum overlap value.

\begin{theorem}
Consider an SC code with parameters $m$, $\kappa$, and $\gamma=2(m+1)$. For a balanced partitioning where there are exactly two elements in any column of $\bold{H}^\textnormal{p}$ that are assigned to each component matrix $\bold{H}_y^\textnormal{p}$, $y\in\{0,\cdots,m\}$, the minimum overlap is:
\begin{equation}
t_\emph{\text{min}}=\left \lceil {\frac{\kappa}{ {{\gamma}\choose{2}} }} \right \rceil.
\end{equation}
\end{theorem}
\begin{IEEEproof}
Assume that the $\kappa$ columns in the protograph matrix $\bold{H}^\textnormal{p}$ are divided into stripes of size $\omega$ columns. Then, the number of stripes is $\left \lceil {\frac{\kappa}{ \omega }} \right \rceil$, and if $\omega$ does not divide $\kappa$, the last stripe has a lower size than $\omega$. Since a stripe is a submatrix in the protograph matrix $\bold{H}$, the overlap parameter is defined for a stripe as well. We denote the overlap parameter for a stripe as $t_{\omega}$. 
The goal is to make each stripe have the minimum overlap parameter value. Each column of $\bold{H}^\textnormal{p}$ has $\gamma=2(m+1)$ elements, and there are two elements in each column that are assigned to the same component matrix. This implies that the minimum possible value for the overlap parameter in each stripe is $t_{\omega}=1$, even if the length of the stripe is $1$ column. Then, we choose the maximum stripe length such that $t_{\omega}=1$ still holds. The maximum $\omega$ such that $t_{\omega}=1$ holds is:
\begin{equation}
\omega_{\text{max}}={{\gamma}\choose{2}},
\end{equation}
which is achieved by choosing a different pair of elements in each column in the stripe to be assigned to the same component matrix. Each stripe adds one unit to the final overlap parameter value. Thus,
\begin{equation}
t_\text{min}=\left \lceil {\frac{\kappa}{ \omega_{\text{max}} }} \right \rceil=\left \lceil {\frac{\kappa}{ {{\gamma}\choose{2}} }} \right \rceil.\vspace{-0.3cm}
\end{equation}
\end{IEEEproof}

Next, we explain a systematic approach to find a partitioning that achieves the minimum overlap parameter. Given one stripe in the protograph matrix $\bold{H}^\textnormal{p}$, the corresponding stripes can be identified for $\bold{H}^\textnormal{p}_y$, $0 \leq y \leq m$, and for any linear combination of them.
In the construction procedure, we focus on properly constructing the first stripe with the overlap parameter $t_{\omega}=1$ in the component matrix $\bold{H}^\textnormal{p}$. The other stripes, except for the last stripe, are constructed by arbitrary permutations of the columns of the first stripe, and the last strip is constructed from an arbitrary subset of columns of the first stripe.
Consider the first stripe, and let $\mathcal{O}^\ell_y$, $0 \leq y \leq m$, $0 \leq \ell \leq {{\gamma}\choose{2}}-1$, be the set of indices of the rows with non-zero values in column $\ell$ of that stripe in $\bold{H}^\textnormal{p}_y$. An element of $\mathcal{O}^\ell_y$ takes a value in $\{0, 1, \dots, \gamma-1\}$. Note that $\vert \mathcal{O}^\ell_y \vert = 2$, $\forall \ell$ and $\forall y$, because our partitioning is balanced.

For the case of $\gamma=4$, the size of each stripe, except for the last stripe, is $\binom{4}{2}=6$ columns. The construction that achieves $t_{\omega}=1$ is simply to select distinct sets $\mathcal{O}^\ell_0$, $\forall \ell$, because this implies that the sets $\mathcal{O}^\ell_1=\{0,\cdots,\gamma-1\}\setminus\mathcal{O}^\ell_0$, for that stripe in $\bold{H}^\textnormal{p}_1$ are also distinct.

For the case of $\gamma=6$, the size of each stripe, except for the last stripe, is $\binom{6}{2}=15$. Here, we develop an algorithm to perform the construction. Let $\mathcal{O}^\ell_{0+1}$, $0 \leq \ell \leq 14$, be the set of indices of the rows with non-zero values in column $\ell$ of our stripe in the matrix $\bold{H}^\textnormal{p}_0+\bold{H}^\textnormal{p}_1$. Note that $\vert \mathcal{O}^\ell_{0+1} \vert = 4$, $\forall \ell$. The algorithm determines each set $\mathcal{O}^\ell_{0+1}$ in the stripe sequentially starting from $\ell=0$. Once the set $\mathcal{O}^\ell_{0+1}$ is determined, the algorithm determines the set $\mathcal{O}^\ell_{0}$, and consequently the set $\mathcal{O}^\ell_{1}$, such that:
\begin{equation}
\mathcal{O}^\ell_{0+1}  \neq \mathcal{O}^r_{0+1}, \text{ }\mathcal{O}^\ell_{0}  \neq \mathcal{O}^r_{0}, \text{ and } \mathcal{O}^\ell_{1}  \neq \mathcal{O}^r_{1}, \text{ } \forall r < \ell.
\end{equation}
There are $\binom{4}{2}=6$ options to choose the set $\mathcal{O}^\ell_{0}$ out of the elements in the set $\mathcal{O}^\ell_{0+1}$. However, not all of these options are valid options because of the constraint in (4). For each $\ell$, the algorithm chooses the set $\mathcal{O}^\ell_{0+1}$ out of $\binom{6}{4}-\ell=15-\ell$ options such that the number of valid options for the set $\mathcal{O}^\ell_{0}$ is maximized. This algorithm always starts with $6$ options to choose the set $\mathcal{O}^0_{0}$ from the elements in $\mathcal{O}^0_{0+1}$, and ends with $1$ option to choose the set $\mathcal{O}^{14}_{0}$ from the elements in $\mathcal{O}^{14}_{0+1}$. Note that once the sets $\mathcal{O}^\ell_{0+1}$, $\forall \ell$, are properly selected, the sets $\mathcal{O}^\ell_{2}$, $\forall \ell$, are properly selected. Consequently, $t_{\text{min}}$ is achieved.

\begin{example}
Let $\gamma=4$, $\kappa=12$, and $m=1$. The length of each stripe is $\omega_\text{max}={{4}\choose{2}}=6$, and the number of stripes is $\kappa/\omega_\text{max}=2$. As a result, $t_\emph{\text{min}}=2$, and a partitioning matrix that achieves the minimum overlap is illustrated below.
\begin{equation*}
\bold{PM}=
\begin{tabular}{|c|c|c|c|c|c|c|c|c|c|c|c|}
\hline
$0$&$0$&$1$&$0$&$1$&$1$&$0$&$0$&$1$&$0$&$1$&$1$\\
\hline
$0$&$1$&$0$&$1$&$0$&$1$&$0$&$1$&$0$&$1$&$0$&$1$\\
\hline
$1$&$0$&$0$&$1$&$1$&$0$&$1$&$0$&$0$&$1$&$1$&$0$\\
\hline
$1$&$1$&$1$&$0$&$0$&$0$&$1$&$1$&$1$&$0$&$0$&$0$\\
\hline
\end{tabular}.
\end{equation*}
For the first stripe (the first six columns in $\bold{H}^\textnormal{p}$), the parameters $\mathcal{O}^\ell_y$, $0 \leq y \leq 1$, $0 \leq \ell \leq 5$, are listed below. The second stripe can be any permutation of the columns of the first stripe. In this example, we chose exactly the same order of columns for the second stripe.
\begin{flalign*}
\begin{split}
\mathcal{O}_0^0=\{0,1\}\hspace{1cm}&\mathcal{O}_1^0=\{2,3\}\\
\mathcal{O}_0^1=\{0,2\}\hspace{1cm}&\mathcal{O}_1^1=\{1,3\}\\
\mathcal{O}_0^2=\{1,2\}\hspace{1cm}&\mathcal{O}_1^2=\{0,3\}\\
\mathcal{O}_0^3=\{0,3\}\hspace{1cm}&\mathcal{O}_1^3=\{1,2\}\\
\mathcal{O}_0^4=\{1,3\}\hspace{1cm}&\mathcal{O}_1^4=\{0,2\}\\
\mathcal{O}_0^5=\{2,3\}\hspace{1cm}&\mathcal{O}_1^5=\{0,1\}
\end{split}
\end{flalign*}
\end{example}
As we noted, the partitioning that achieves the minimum overlap value is not unique. We suggest two methods for choosing the MO partitioning and constructing the final SC code. The first method is searching over all MO partitioning choices and finding the one that results in the minimum number of problematic objects in the lifted graph for a fixed arrangement of the circulant powers \cite{MOP_ISIT_2017}. This technique is not practical for high values of the row or column weights because of the dramatic number of partitioning that achieves the minimum overlap parameter value.%As a result, searching among all the options brings a complexity complexity issue. 

The second method is choosing one MO partitioning and an initial set of circulant powers. Then, we apply the CPO algorithm to reduce the population of problematic objects by adjusting the circulant powers. The logic behind the second method is that all MO partitioning choices for the case $\gamma=2(m+1)$ result in the protograph matrices with similar number of problematic objects, and the difference, which appears after lifting, can be compensated by applying the CPO algorithm.

\section{SC Code Design to Alleviate SNR Variation}
In this section, we first describe our framework to construct SC codes over channels with SNR variation, and explain why well-designed SC codes show a superior performance compared to block codes with interleaving. Then, we study the effect of parameter memory on the performance of SC codes over channels with SNR variation. Finally, we introduce an interleaving scheme for SC codes that further improves their performance for channels with non-uniform SNR. \vspace{-0.2cm}

\subsection{Code Design Machinery}

Instead of using an individual block code for each section of the channel in Fig.~3(a), we use an SC code that spans several consecutive sections. By using SC codes, we can make more reliable sections help unreliable ones while keeping the decoding latency low. The decoding latency of an SC code is a function of the underlying code length and the size of sliding window of the decoder \cite{6374679}.

Parameters of our code design are illustrated in Fig.~4. The length of the underlying block code is equal to the length of one section of the channel, thus each replica of an SC code spans one section of the channel. The coupling length $L$ determines how many sections are spanned by one SC codeword, i.e., $N=L$. The parameter $L$ must be chosen such that a variety of sections with different reliabilities are included. We use CB codes as underlying block codes. For partitioning the underlying block code, we use the MO partitioning scheme.
%The MO partitioning aims to minimize the number of combinatorial structures in the Tanner graph of SC codes that are problematic over AWGN channel.
In our channel model, each section of the channel is modeled as an AWGN channel, thus the MO partitioning technique is a suitable choice \cite{MOP_ISIT_2017}.

\begin{figure}
\centering
\includegraphics[width=0.35\textwidth]{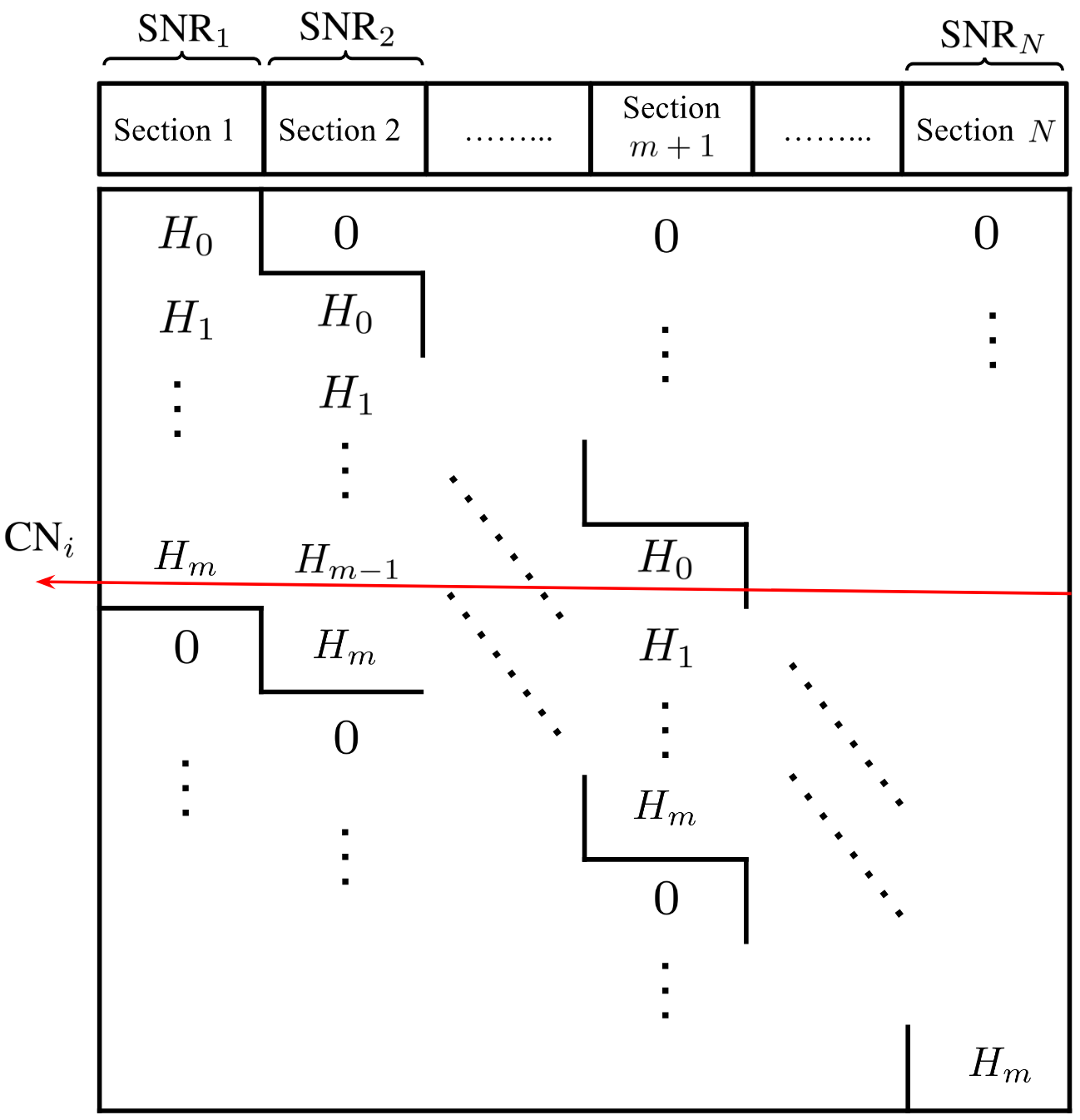}\\
\caption{An SC code with memory $m$ constructed for a non-uniform channel, $\text{CN}_i$ spans ($m+1$) consecutive sections.\vspace{-0.3cm}}
\end{figure}

The memory $m$ of an SC code plays a critical role in the performance over channels with SNR variation.  The parameter $m$ determines how many different sections the VNs of a check equation span. All CNs of an SC code receive messages from VNs within ($m+1$) consecutive sections (see $\text{CN}_i$ in Fig.~4), except for the first and last group of $m\gamma p$ CNs which receive messages from a fewer number of sections. 
{As a result, if a CN is connected to an unreliable VN, the other more reliable neighboring VNs of that CN can help correcting the VN error.}
%As a result, if there is one reliable section with a high SNR, it can help messages from the other $m$ sections be recovered.
If $m$ is chosen appropriately, each CN receives enough reliable messages to send improved messages to its neighboring VNs in an iterative decoding. However, there is a trade-off, and as we increase the memory, the decoding window size and consequently, the decoding latency also increase.

The simulation results in Section V show that our SC codes outperform block codes (with interleaving) for channels with SNR variation. We also demonstrate that the performance gap over channels with SNR variation and corresponding uniform channels is lower for SC codes compared to block codes. This result verifies that the correlation that exists between different sections due to the structure of SC codes causes enough diversity to alleviate the negative impact of channel non-uniformity.  \vspace{-0.4cm}

\subsection{Interleaving for SC Codes}

In this section, we introduce a low-cost interleaving scheme specialized for SC codes in order to further improve the performance. In the code design approach described in the previous subsection, CNs receive messages from ($m+1$) different sections. If most of these ($m+1$) sections have a low SNR, the decoder may fail to recover the message correctly. We use interleaving to avoid letting multiple unreliable VNs from consecutive sections participate in one check equation and degrade the performance.

Consider an  SC code with the coupling length $L$ ($L$ is also equal to the number of sections covered by one SC codeword) and memory $m$. We assume that ($m+1$) divides $L$. We divide the SC codeword into $L$ groups, and we further divide each group of data into $L/(m+1)$ chunks. Then, we rearrange them by taking one chunk from each group in order and placing them next to each other, see Fig.~5. This interleaved data is passed through the channel, and the de-interleaving is performed on the received data from the channel and before decoding. Due to interleaving, each CN receives equal number of messages from all $L$ different levels of reliabilities (except for the first and last groups of CNs).

\begin{figure}
\centering
\includegraphics[width=0.45\textwidth]{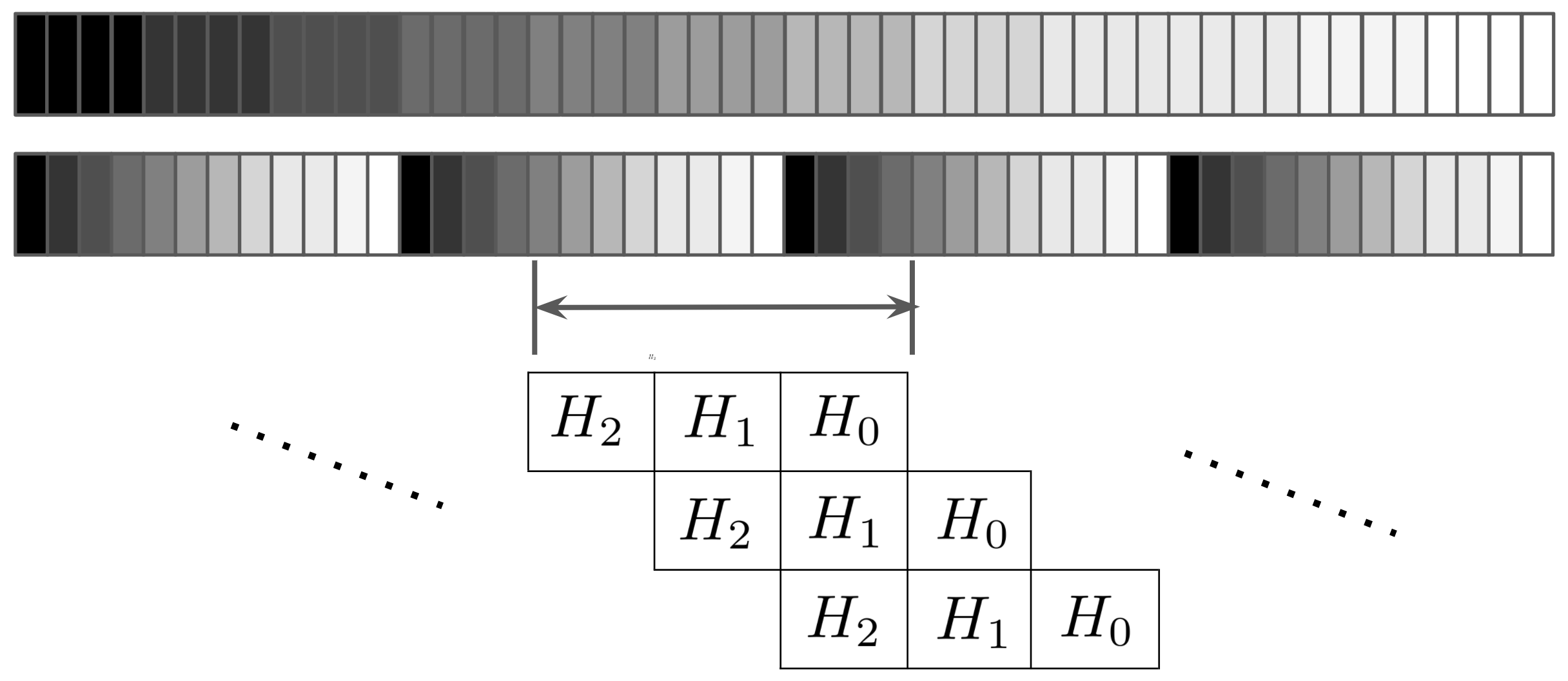}\\
\caption{The original sequence of data (top panel) represents the worst case scenario where multiple consecutive replicas are affected by a low SNR. Interleaved sequence of data (middle panel) for an SC code with $m=2$ and $L=12$: a darker gray represents a lower SNR. These chunks are interleaved such that each check equation receives messages with all $L$ different reliabilities.\vspace{-0.5cm}}
\end{figure}
\begin{claim}
Consider an SC code with parameters $m$ and $L$. Let $\mu$ be the number of chunks that an SC codeword from this code is partitioned into for the interleaving step. Then, the smallest value of $\mu$ that provides the same average reliability for all CNs, except for the first and last few CNs that have smaller degrees, is $\mu_\emph{\text{opt}}=L^2/(m+1)$.
\end{claim}
\begin{IEEEproof}
We prove Claim~1 by the contradiction. Suppose an SC codeword from this code is partitioned into $\mu<L^2/(m+1)$ chunks. 
In an iterative decoding of an SC codeword, except for the first and last few CNs, all CNs receive messages from a contiguous $(m+1)/L$ fraction of all VNs. In other words, they receive messages from $\nu=\mu(m+1)/L<L$ consecutive chunks. Suppose that the interleaving is perfect in the sense that all these $\nu$ chunks have different SNRs. Then, the number of different SNRs that a CN is affected by, $\nu$, is less than the total number of different SNRs, $L$. Consequently, some CNs may receive messages from more reliable VNs while others do not, and the average SNR need not be the same for check equations.
\end{IEEEproof}

Our presented interleaving scheme partitions an SC codeword into $\mu_\text{opt}=L^2/(m+1)$ chunks, and ensures that all check equations, except for the first and last few ones, have the same average SNR. 
For example, an SC codeword with $m=2$ and $L=12$, shown in Fig.~5, is partitioned into $\mu_\text{opt}=48$ chunks. By using interleaving, we avoid letting multiple unreliable sections dominate the messages received by one CN. Our simulation results show that interleaving notably reduces the performance gap that exists between the error rates of SC codes over non-uniform and uniform channels.
{We note that using block codes and regular interleaving requires {dividing} the data with the same length into $L^2$ {chunks}.}

We note that the performance of the SC code constructed using our technique is not sensitive to the perfect alignment of underlying codes and sections of the channel. In fact, a substantial mis-alignment resembles an uninformed interleaving that neither helps nor degrades the performance.\vspace{-0.4cm}

\section{Simulation Results}

\begin{figure*}
\tiny
\centering
\begin{tabular}{cc}
\begin{tabular}{|c|c|c|c|c|c|c|c|c|c|c|c|c|c|c|c|c|}
\hline
0&0&1&1&0&1&1&1&0&0&0&0&0&1&1&1&1\\
\hline
1&1&0&0&1&0&1&1&0&1&1&1&0&0&1&0&0\\
\hline
1&0&0&0&1&0&0&0&1&1&0&0&1&1&0&1&1\\
\hline
\end{tabular}
&
\begin{tabular}{|c|c|c|c|c|c|c|c|c|c|c|c|c|c|c|c|c|}
\hline
0&1&0&2&2&1&2&0&0&1&0&1&1&2&1&2&2\\
\hline
1&0&2&1&0&2&1&1&1&2&2&0&2&0&0&0&1\\
\hline
2&2&1&0&1&0&0&2&2&0&1&2&0&1&2&1&0\\
\hline
\end{tabular}\vspace{0.15cm}\\
(a)&(b)\vspace{0.15cm}\\
\multicolumn{2}{c}{
\begin{tabular}{|c|c|c|c|c|c|c|c|c|c|c|c|c|c|c|c|c|c|c|c|c|c|c|c|c|c|c|c|c|}
\hline
1&1&2&1&0&1&2&2&0&2&1&0&2&0&0&1&1&2&1&0&1&2&2&0&2&1&0&2&0\\
\hline
1&2&1&0&1&2&1&0&2&2&0&2&0&1&0&1&2&1&0&1&2&1&0&2&2&0&2&0&1\\
\hline
2&1&0&2&1&2&0&2&1&0&2&1&0&0&1&2&1&0&2&1&2&0&2&1&0&2&1&0&0\\
\hline
2&0&1&1&2&0&2&1&2&0&0&0&1&2&1&2&0&1&1&2&0&2&1&2&0&0&0&1&2\\ 
\hline
0&2&2&2&2&0&0&0&0&1&1&1&1&1&2&0&2&2&2&2&0&0&0&0&1&1&1&1&1\\ 
\hline
0&0&0&0&0&1&1&1&1&1&2&2&2&2&2&0&0&0&0&0&1&1&1&1&1&2&2&2&2\\
\hline
\end{tabular}\vspace{0.15cm}}\\
\multicolumn{2}{c}{(c)}\vspace{0.15cm}\\
\multicolumn{2}{c}{
\begin{tabular}{|c|c|c|c|c|c|c|c|c|c|c|c|c|c|c|c|c|c|c|c|c|c|c|c|c|c|c|c|c|}
\hline
4&55&59&49&10&4&59&14&17&44&12&58&55&27&11&15&16&15& 7& 0&36& 0& 0&22&41&0&0&0&24\\
\hline
20&28&28&3&15&5&54&7&8&9&49&11&12&13&14&15&16&59&18&52&30&21&22&23&24&25&26&50&34\\
\hline
0&14&4&6&8&10&11&14&16&18&20&22&12&19&32&30&1&34&36&38&40&42&44&46&48&42&26&54&56\\
\hline
8&3&6&9&12&15&57&32&24&27&24&55&57&39&42&45&48&51&17&57&60&2&5&8&11&14&17& 20& 23\\
\hline
0 &4&8&8&16&20&24&28&32&4&40&44&18&52&56&60&27&7&11&15&19&23&27&31&17&39&5&47&51\\
\hline
0&5&10&17&20&25&30&37&40&13&50&55&57&4&3&14&19&36&49&28&20&47&49&31&59&3&8&13&18\\
\hline
\end{tabular}\vspace{0.15cm}}\\
\multicolumn{2}{c}{(d)}\vspace{0.15cm}
\end{tabular}\vspace{-0.15cm}
\caption{(a) Partitioning matrix for SC~Code~1. (b) Partitioning matrix for SC~Code~2. (c) Partitioning matrix for SC~Code~3. (d) Circulant power matrix for SC~Code~3.\vspace{-0.05cm}}
\end{figure*}

In this section, we present our simulation results. First, we show the performance gap that exists over channels with uniform and non-uniform SNRs for uncoupled individual block codes. Next, we compare the performance of SC codes possessing the introduced structure with block codes, and illustrate the effect of interleaving and increasing the memory.
We also demonstrate the outstanding performance of SC codes with column weight $\gamma=2(m+1)$. The MO partitioning for $\gamma=4$ was discussed in Example~2. In the simulation results, we consider SC codes with $\gamma=6$ constructed by the MO partitioning technique and CPO algorithm, over uniform and non-uniform channels.
All simulations were performed using a high performance platform with a lot of computational cores. We used a min-sum algorithm with a maximum of $50$ iterations for decoding.

Our code parameters are as follows. Block~Code~1 is a CB block code with $\kappa=z=17$, $\gamma=3$, and $f_{i,j}=ij$. Block~Code~1 has length $289$ bits and rate $r_\textnormal{d}\approx0.824$, and it is the underlying block code for SC~Codes~1 and 2. SC~Codes~1 and 2 are constructed by MO partitioning \cite{MOP_ISIT_2017}, and they have coupling length $L=30$ and length $8670$ bits. The memory and rate for SC~Code~1 are $m=1$ and $r_\textnormal{d}\approx0.818$, and for SC~Code~2 are $m=2$ and $r_\textnormal{d}\approx0.812$. Figs.~6(a)~and~6(b) show the partitioning matrices for SC~Codes~1 and 2, respectively. %For the block codes, we use regular interleaving reviewed in Section~II, and for SC codes, we use the cheaper interleaving scheme introduced in Section~III. 

SC~Code~3 is an SC code with $\kappa=29$, $\gamma=6$, $z=61$, $m=2$, and $L=6$. It is constructed by the extended MO partitioning, and its circulant powers are optimized using the CPO algorithm. According to Theorem~1, the length of each stripe (except for the last stripe) is $\omega_\text{max}={{6}\choose{2}}=15$, and the number of stripes is $\left \lceil \frac{\kappa}{\omega_\text{max}} \right \rceil=2$. As a result, $t_\text{min}=2$, and the partitioning matrix for SC~Code~3 that achieves the minimum overlap is illustrated in Fig.~6(c). Let $\bold{CP}=[f_{i,j}]$ be the circulant power matrix with dimension $\gamma\times\kappa$. The circulant power matrix of SC~Code~3 after applying the CPO algorithm is depicted in Fig.~6(d).

For Block~Code~1 and SC~Code~1 and 2, we use a non-uniform channel with $N=30$ sections. For SC~Code~3, we use a non-uniform channel with $N=6$ sections. For block codes, we use the interleaving technique that is reviewed in the preliminaries, and for SC codes, we use the interleaving scheme introduced in this paper. In a practical channel model, the values of $\Delta \text{SNR}_{s}$ for different sections are dependent. We describe the correlation model among SNRs of different sections as follows. Let $\textbf{a}$ be an $\alpha\times 1$ vector that describes the correlation coefficients among $\alpha$ consecutive sections:
\vspace{-0.1cm}
\begin{align}
&{(\Delta \text{SNR}_{s})}_\text{dB}=\textbf{a}^\textnormal{T}\textbf{u}, \nonumber \\
&\textbf{u}=[u_s\text{ }u_{s-1}\text{ }\cdots\text{ }u_{s-\alpha+1}]^\textnormal{T},\hspace{0.1cm}u_i\sim\mathcal{N}(0,\sigma^2).
\end{align}
$\mathcal{N}(0,\sigma^2)$ defines a Gaussian distribution with mean $0$ and variance $\sigma^2$. In our simulations, we consider the following parameters for the channel:\footnote{On a hard disk drive, under certain environmental and vibration conditions, BER for each of many consecutive sectors around a circular track can be measured. The autocorrelation terms gauge the coupling of SNR between sectors of various spacing. The autocorrelation model used in this paper come from experimental data provided by a Western Digital Company (WDC).}
\begin{align}
\textbf{a}_{\text{ }}{=}&[0.78\text{ }0.44\text{ }0.31\text{ }0.23\text{ }0.15\text{ }0.08\text{ }0.09\text{ }0.03\text{ }0.04\text{ }0.02 \nonumber \\
&{-}0.04]^\textnormal{T},\hspace{0.2cm}\alpha_{\text{ }}{=}\text{ }11,\hspace{0.1cm}\sigma{=}\text{ }0.15.
\end{align}

Fig.~7 shows the BER curves for Block~Code~1 over non-uniform and uniform channels with the same average SNR. Because of the SNR variation, the performance over non-uniform channel is around $1$ order of magnitude worse in the error floor region. Then, we apply regular interleaving for the non-uniform channel. It can be seen, as expected, that interleaving compensates for the performance loss due to the SNR variation. (For the non-uniform channel, the horizontal axis represents ${\text{SNR}_{\text{abs}}}$.)

\begin{figure}
\centering
\vspace{-0.8cm}\includegraphics[width=0.45\textwidth]{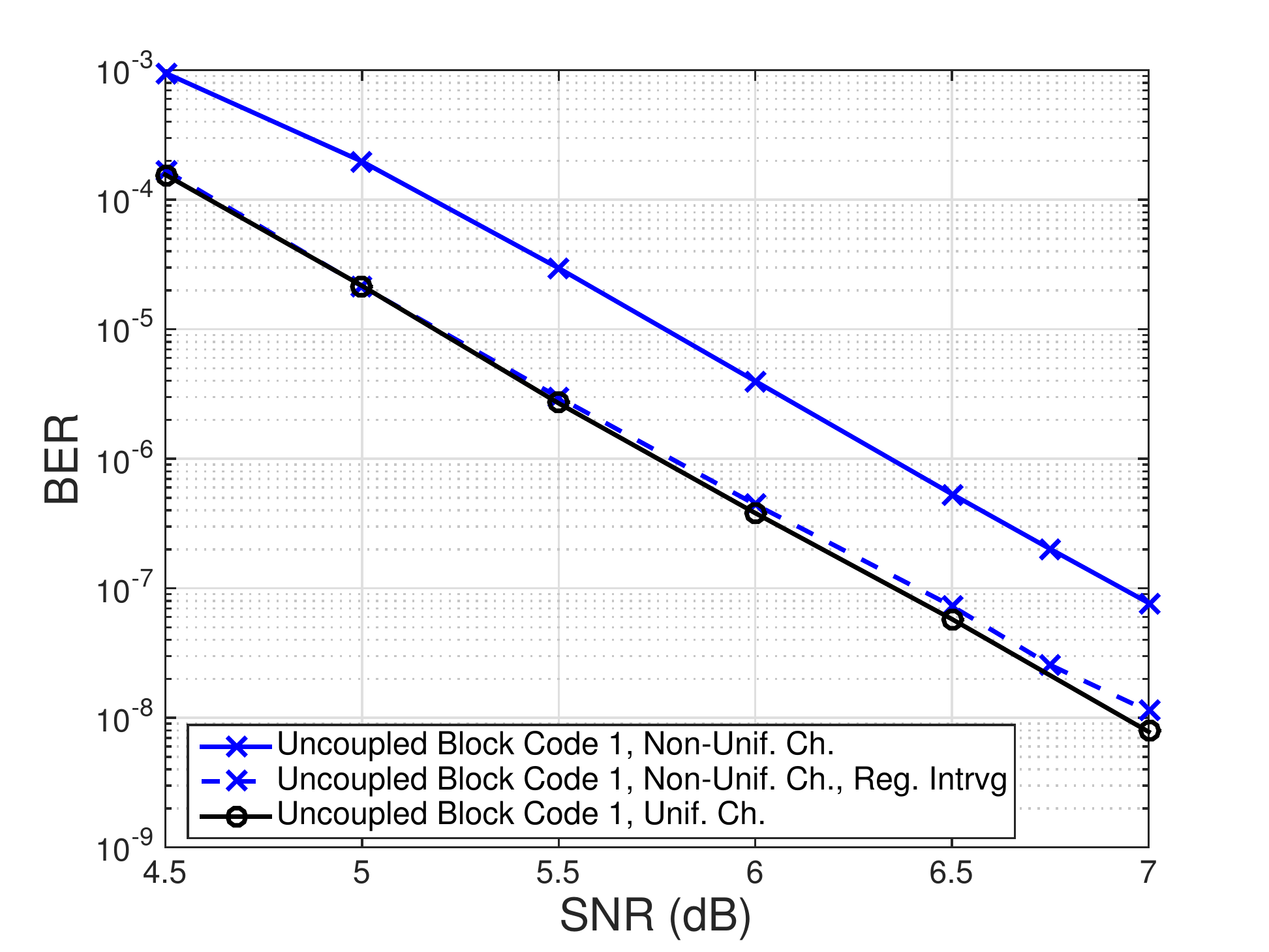}\\
\caption{BER curves for Block~Code~1 over uniform and non-uniform channels with/without interleaving.}
\vspace{-0.2cm}
\end{figure}

Next, we compare the error floor performance of uncoupled block codes with SC codes over channels with SNR variation. According to Fig.~8: 1) SC~Code~1 shows $1$ and $2$ orders of magnitude performance improvement compared to the uncoupled chain of Block~Code~1 with and without interleaving, respectively, while it has comparable decoding latency, and 2) SC~Code~2 secures even further improvement by providing more diversity.

\begin{figure}
\centering
\vspace{-0.35cm}\includegraphics[width=0.45\textwidth]{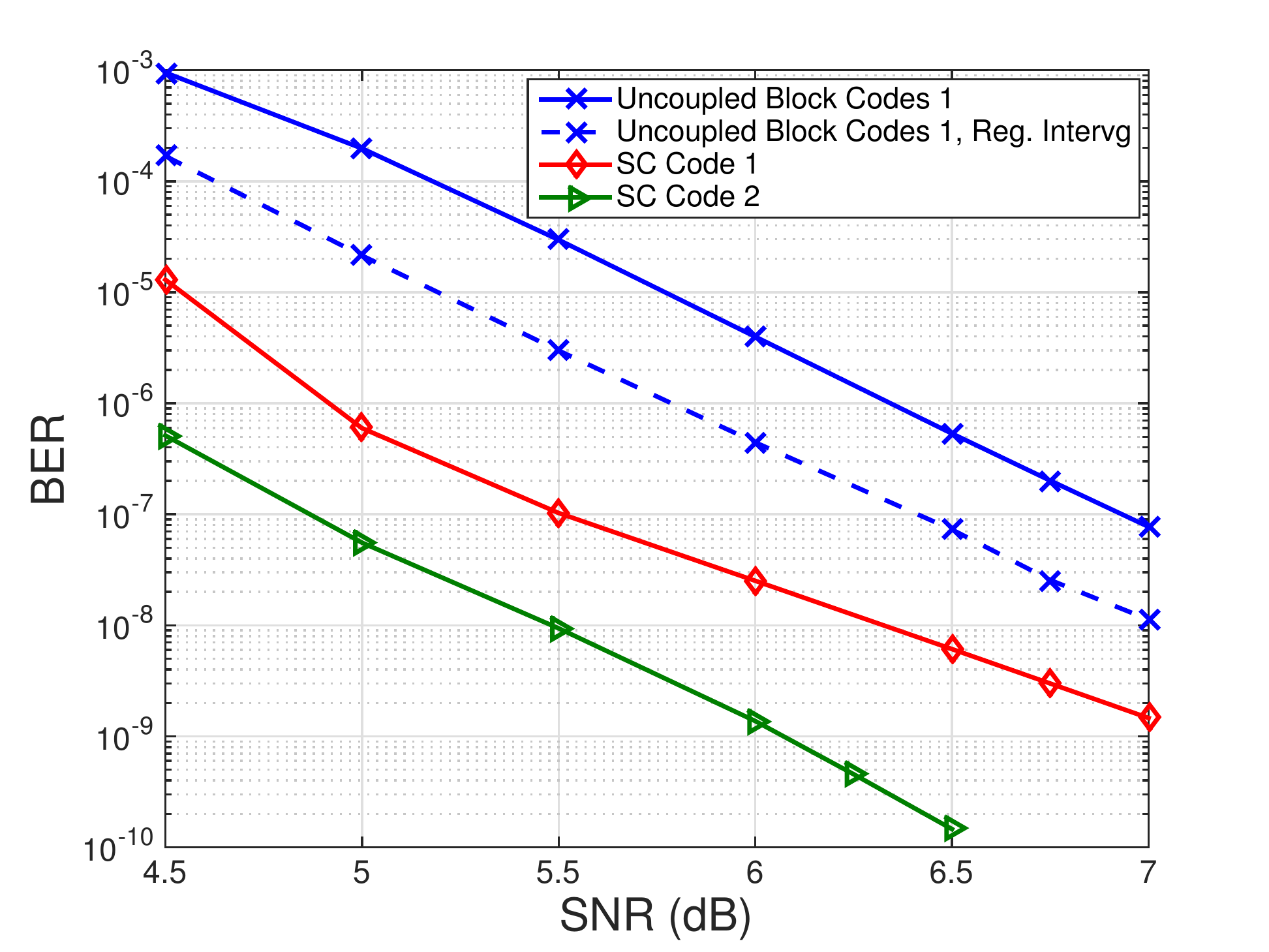}\\
\caption{BER curves for Block code with/without interleaving vs. SC code with memories $m=1$ and $2$  over non-uniform channel.\vspace{-0.5cm}}
\end{figure}

Fig.~9 demonstrates the performance loss due to SNR variation for SC~Code~1. The performance loss at SNR$=6.5$~dB is around $0.5$ of an order of magnitude. Compared to the performance loss for Block~Code~1, we immediately see that well-designed SC codes are more robust against SNR variation. Furthermore, this loss can be compensated by the efficient interleaving introduced in this paper, as Fig.~9 shows. 

\begin{figure}
\centering
\vspace{-0.8cm}\includegraphics[width=0.45\textwidth]{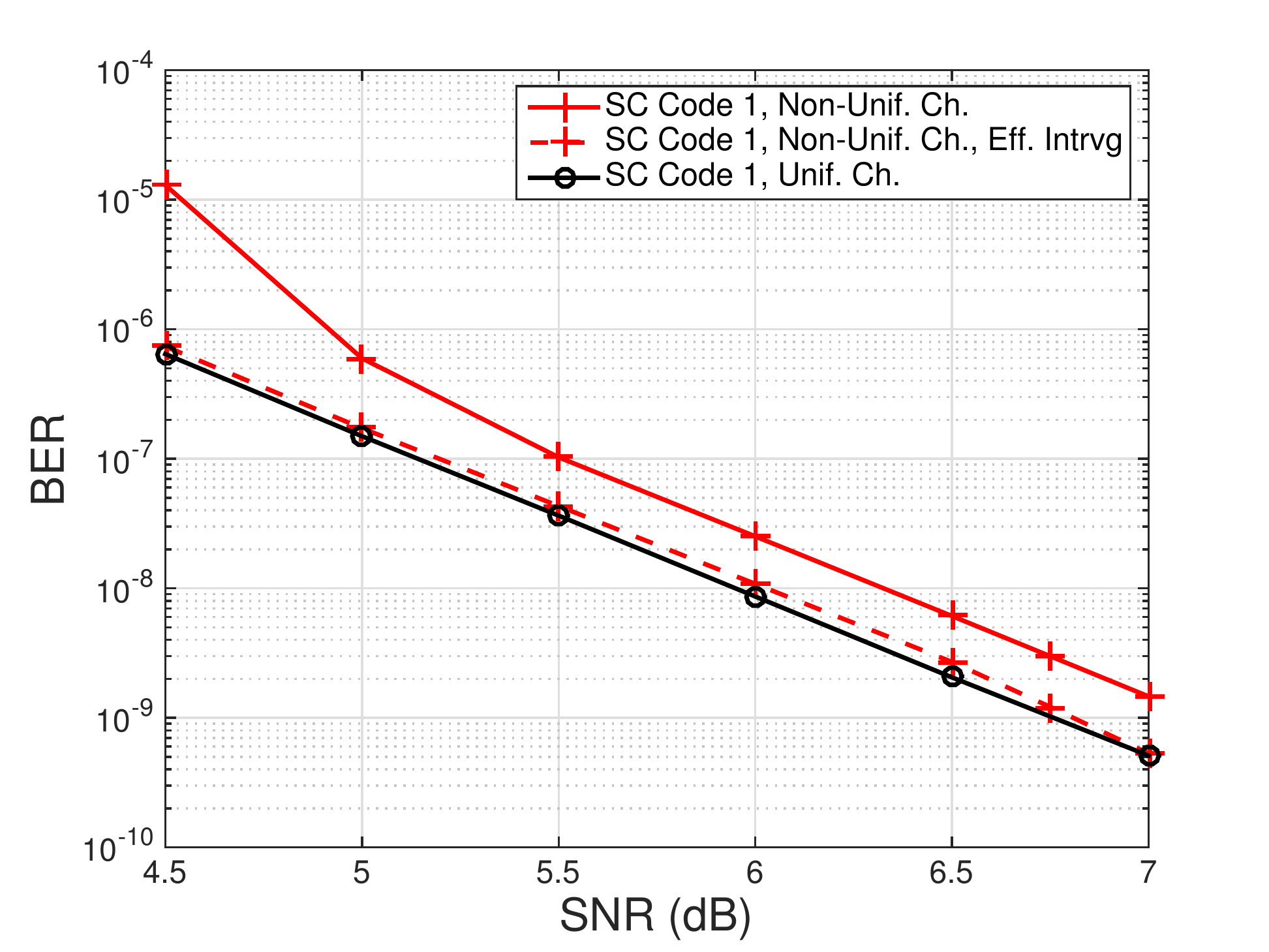}\\
\caption{BER curves for SC~Code~1 over uniform and non-uniform channels with/without interleaving.
\vspace{-0.23cm}}
\end{figure}

Fig.~10 demonstrates the similar analysis for SC~Code~3.
Since well-designed SC codes with $\gamma=6$ have a very low error floor, we could not collect enough errors in the high SNR region. As we see in this figure, our well-designed SC code with column weight $\gamma=6$ constructed by the extended MO partitioning has a sharp waterfall region, and it achieves bit error rate (BER) $10^{-10}$ at SNR$=4.1$~dB. We also note that there is a notable performance gap between uniform and non-uniform channels for this code which is because of the high SNR variation of the channel with $N=6$ sections and the relatively low coupling length of the SC code ($L=6$). However, our efficient interleaving scheme remarkably reduces this performance gap.

\begin{figure}
\centering
\vspace{-0.35cm}\includegraphics[width=0.45\textwidth]{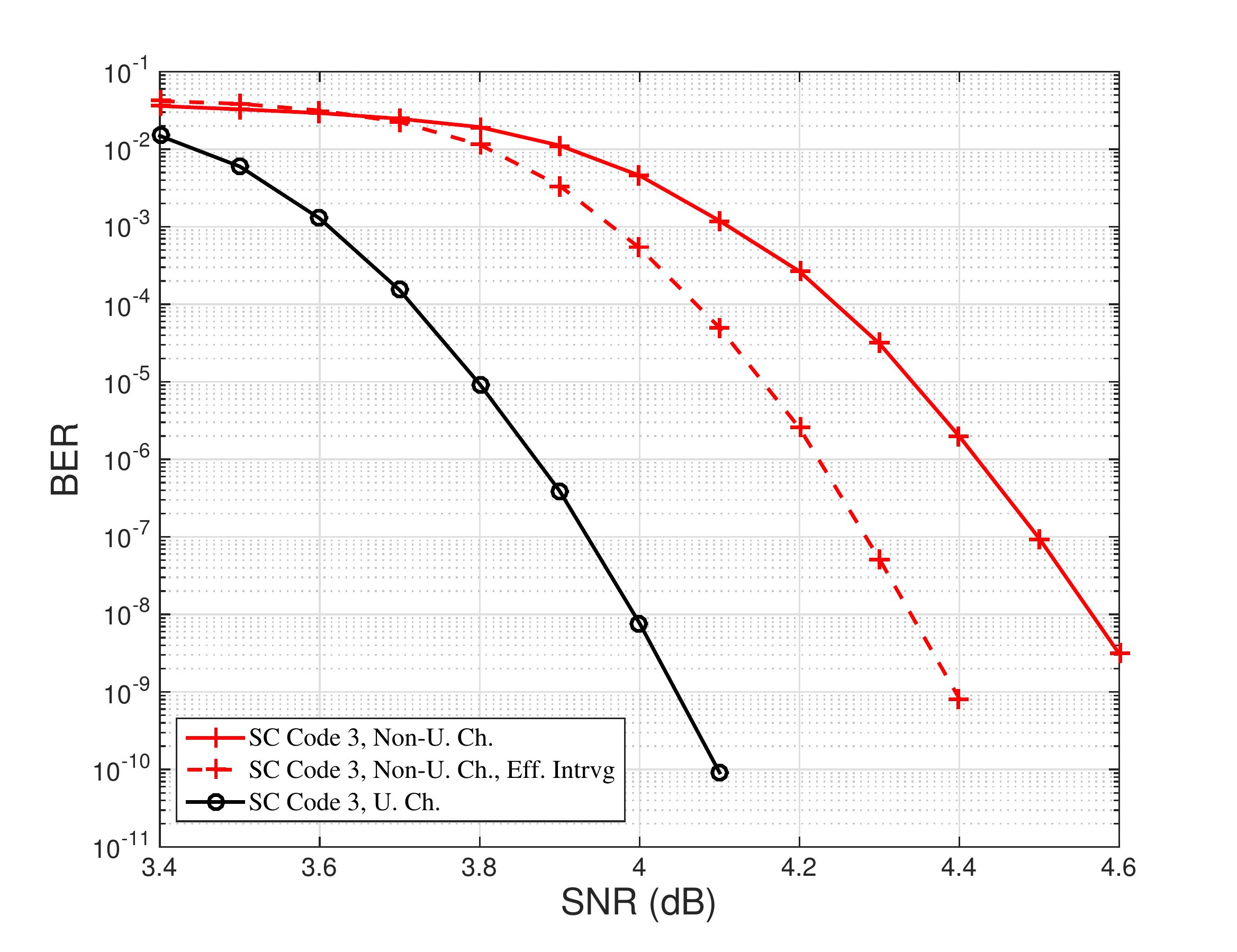}\\
\caption{BER curves for SC~Code~3 over uniform and non-uniform channels with/without interleaving.\vspace{-0.5cm}}
\end{figure}

We have also performed a comparison between uncoupled Block~Codes~1 with regular interleaving and the SC~Code~1 with our proposed interleaving, with respect to the average number of iterations the decoder needs to converge at different absolute SNRs. We consider a number of sectors $N=30$ and thus, $L=30$. The uncoupled block codewords are divided into groups of $30$ codewords each. Convergence of the decoder is then achieved only if all the $30$ transmitted blocks in the group are decoded. Mathematically, let $M_{\textup{BL}}$ be the number of uncoupled block codewords simulated (encoded, transmitted, then decoded), then $M_{\textup{BL}}/30$ is the number of groups ($30$ divides $M_{\textup{BL}}$). Moreover, let the pair $(\beta_1, \beta_2)$, $1 \leq \beta_1 \leq M_{\textup{BL}}/30$ and $1 \leq \beta_2 \leq 30$, be the group index and the block codeword index within the group, respectively. Thus, the average number of iterations in the case of the uncoupled block codes is:
\begin{equation}
\mathcal{I}_{BL} = \frac{30}{M_{\textup{BL}}} \sum_{\beta_1=1}^{M_{\textup{BL}}/30} \max_{\beta_2 \in \{1, 2, \dots, 30\}} \mathcal{I}_{\beta_1, \beta_2},
\end{equation}
where $\mathcal{I}_{\beta_1, \beta_2}$ is the number of iterations required for the block codeword indexed by $(\beta_1, \beta_2)$. Similarly, let $M_{\textup{SC}}$ be the number of SC codewords simulated. Moreover, let $\beta_3$, $1 \leq \beta_3 \leq M_{\textup{SC}}$, be the SC codeword index. Thus, the average number of iterations in the case of the SC code is:
\begin{equation}
\mathcal{I}_{SC} = \frac{1}{N_{\textup{SC}}} \sum_{\beta_3=1}^{N_{\textup{SC}}} \mathcal{J}_{\beta_3},
\end{equation}
where $\mathcal{J}_{\beta_3}$ is the number of iterations required for the SC codeword indexed by $\beta_3$.

Our simulation results demonstrate that the average number of iterations needed in the case of the SC code is consistently less than the average number of iterations needed in the case of the uncoupled block codes for all the absolute SNRs we did this comparison at. Furthermore, increasing the absolute SNR reduces the difference between the two average numbers of iterations. This observation is justified by the fact that the higher the absolute SNR is, the less the number of errors and their magnitudes are, i.e., the effect of the performance of the code on the number of iterations needed for convergence becomes less. In Fig. 11, these average numbers of iterations are plotted versus the absolute SNR.\vspace{-0.2cm}

\begin{figure}
\centering
\vspace{-0.4cm}
\includegraphics[width=0.45\textwidth]{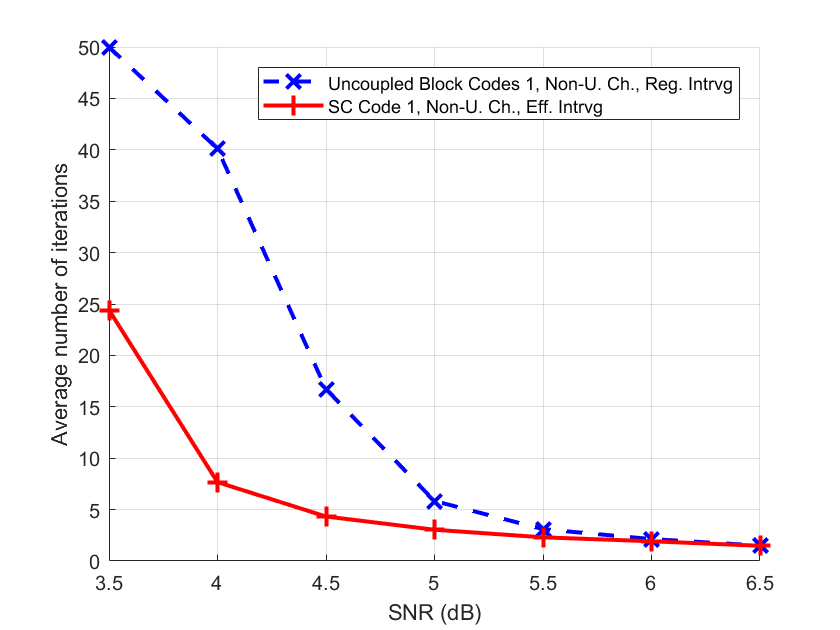}
\caption{Average number of iterations to converge for uncoupled block codes vs. SC code.\vspace{-0.5cm}}
\end{figure}

\section{Conclusion}
In this paper, we presented a specially-coupled (SC) code design for channels with SNR variation. First, we introduced a new construction method for SC codes with higher column weights. Second, we demonstrated that our well-designed SC codes outperform block codes thanks to the diversity that is provided by the coupling, while they have comparable latency as their underlying block codes.  Finally, we also demonstrated that the performance of SC codes can be further improved by increasing the memory and performing the interleaving scheme introduced in this paper.  {One of our future works is to investigate the performance of our well-designed SC codes and the proposed efficient interleaving over partial response channels with SNR variation.}\vspace{-0.2cm}

\vspace{-0.01cm}
\section*{Acknowledgments}
\vspace{-0.02cm}
Research supported by a grant form ASTC-IDEMA.

\vspace{-0.06cm}
\balance
\bibliographystyle{IEEEtran}
\bibliography{IEEEabrv,references}

\end{document}